# Allying nanophotonic structures with two-dimensional van der Waals materials

Yuan Meng[1,2 †], Hongkun Zhong[1 †], Zhihao Xu[3 †], Tiantian He[1], Justin S. Kim[3], Sangmoon Han[3], Yijie Shen[4], Mali Gong[1], Sang-Hoon Bae[3,*], Qirong Xiao[1,*]

[1]State Key Laboratory of Precision Measurement Technology and Instruments, Department of Precision Instrument, Tsinghua University, Beijing, China

[2]Present address: Department of Mechanical Engineering and Materials Science, Washington University in St. Louis, St. Louis, MO, USA

[3]Institute of Materials Science and Engineering, Washington University in St. Louis, St. Louis, MO, USA

[4]Optoelectronics Research Centre, University of Southampton, Southampton SO17 1BJ, United Kingdom

[†] Equal contribution. Emails: sbae22@wustl.edu, xiaoqirong@tsinghua.edu.cn

## Abstract

The integration of two-dimensional (2D) materials with photonic structures has catalyzed a wide spectrum of optical and optoelectronic applications. Conventional nanophotonic structures generally lack efficient reconfigurability and multifunctionality. The atomically thin 2D van der Waals materials can thus infuse new functionality and reconfigurability to the well-established library of photonic structures such as integrated waveguides, optical fibers, photonic crystals, micro-cavities, and metasurface, to name a few. Thanks to the handiness of van der Waals interfaces, the 2D materials can be easily transferred and mixed with other prefabricated photonic templates with high degrees of freedom, and can act as the optical gain, modulation, sensing, or plasmonic media for diverse applications. Here we review recent advents on combining 2D materials to nanophotonic structures for new functionality development or performance enhancements. Challenges and emerging opportunities in integrating van der Waals building blocks beyond 2D materials are also discussed.

## 1. Introduction

The integration of functional materials to photonic structures[1] is essential for building high-performance integrated optoelectronic systems[2-4] and an ideal platform for investigating nanophotonic phenomena[5]. The field of photonic integrated circuits[6-8] and nanophotonics[9-11] has achieved vibrant progress for optical communications[12-14], artificial intelligence[15-17], quantum technology[18-20], imaging[21-24], sensing[25,26], and display[27-29], to name a few. However, issues such as insufficient multifunctionality and reconfigurability still await further attention. To address these challenges, novel functional materials are required beyond conventional single silicon or silicon nitride photonics platform[30,31] to infuse new functionalities or impart versatile reconfigurability to the currently established photonic nanostructures[1,32,33].

Two-dimensional (2D) materials are one of such candidates with promising optoelectronic attributes[32,34-36], and have attracted immense research interest over the past decades since the debut of graphene[37]. Including graphene[38], transition-metal dichalcogenides (TMDs)[39], 2D carbides and nitrides (MXenes)[40-42], hexagonal boron nitride (h-BN)[43], and emergent candidates like quasi-2D halide perovskites[44-52], a big family of 2D materials are established, with a vast collection of available bandgap values[53]. As naturally layered materials with interlayer van der Waals interactions, these materials can be exfoliated into atomic monolayers to unveil extraordinary electronic and optoelectronic properties[54].

Conventional approaches to impart functional materials rely on heteroepitaxy, which encounters lattice-matching and process compatibility constraints[55,56], greatly limiting the possible material combos for heterogeneous integration in photonics. When the epilayer and substrate have different crystal structure or same crystal structure but with big lattice parameter distinction, poly-crystalline phase and defects tend to be generated above certain thickness with compromised material performance[57]. In contrast, devoid of the one-to-one chemical bonding between the material layer and substrate, the 2D materials with signature van der Waals interfaces have made them very easy to transfer, mix, and integrate with other material platforms or prefabricated photonic architectures[58]. As prime building blocks for van der Waals integration[1,58-63], the ultrathin and flexible nature of 2D materials also enable the transfer onto three-dimensional (3D) non-planar optical structures and permit wearable, implantable, and bio-compatible optical applications[64-72].

In this Perspectives, we outline recent advances in van der Waals materials-enabled nanophotonic applications[1,73,74]. We highlight the intriguing outlook on combining 2D van der Waals materials and photonic structures for infusing novel device functionality and reconfigurability to conventional photonic structures and the emergent platform to investigate nanophotonic physics[5,75-77]. Fundamental attributes of representative 2D materials and their coupling to exemplary optical structures, such as dielectric waveguides, optical fibers, photonic crystals, and metasurfaces[14,78,79], are cataloged. Furthermore, we also underline recent advents on the van der Waals integration of 3D freestanding nanomembranes. Awaiting challenges and exciting opportunities in this field are also discussed based on current perspectives.

## 2. Fundamentals for 2D photonics

Nanophotonic structures can be functionalized with diverse 2D materials depending on the base function of the optical architectures and the intrinsic attributes of the 2D materials[1,75,80-84]. The valuable fundamental optoelectronic properties and practical considerations on engineering their optical physical coupling are discussed as the following[85].

### 2.1 2D material fundamentals

Frequently applied 2D van der Waals materials in nanophotonics are typically monolayer to few atomic layers in order to bespeak extraordinary attributes[35,38]. Their outstanding properties along with the atomically clean and electronically keen vdW interfaces to build artificial vdW heterostructures[58,86,87] and Moiré superlattices[88-91] have rapidly burgeoned as one of the prominent research directions in material science, photonics, and nanotechnology communities. The complementary metal-oxide semiconductor (CMOS)-compatibility and handiness for transfer have made them candidates for nanophotonic integration[1,92].

As a monolayer sheet comprised of carbon atoms in honeycomb-lattice, graphene is a zero-bandgap semimetal with beneficial low density of states[38], high carrier mobility[93], and tunable optical transitions[94]. Its unique band structure also enables nonlinear optics[95,96] and gate-tunable surface plasmon polaritons (SPPs)[97-99], as long as applications for transparent electrodes[100-103], sensors[104], and broadband optical modulation and photodetection[105].

In contrast to graphene that lacks an electronic bandgap, 2D semiconductors such as the black phosphorous with puckered atomic monolayers exhibit a broadly tunable bandgap from around 0.3 eV to 2.0 eV depending on layer numbers from bulk to monolayer[106-108]. For TMD with general formula of $MX_2$, where "M" represents transition-metal atom (such as Mo and W), and "X" stands for chalcogen atom (such as S, Se, or Te)[35], their bandgap changes from indirect to direct when exfoliating to monolayers[53]. Therefore, these 2D semiconductors are ideal candidates to realize active optical applications, such as light sources, optical amplification, transistors, neuromorphic computing units, and photodetection[109-118]. Moreover, a majority of 2D TMDs also possess alluring exitonic and polaritonic physical attributes that are worth studying[90,119-123].

h-BN is wide-bandgap (~ 6eV) 2D material that was originally applied as the optimal substrate for graphene and can be used as insulating or encapsulation layers[43,124]. It can provide dangling-bond-free interfaces for photonic van der Waals integration[1], possessing also hyperbolic phonon polaritons[125-127], single-photon emission[128], and second-order nonlinearity[129].

Besides conventional 2D materials, 2D MXenes refer to a set of transition metal carbides, carbonitrides and nitrides with layered lattices bounded by van der Waals forces as emergent 2D candidates[40,130-132]. With the general chemical formula of $M_{n+1}X_nT_x$ ($n$ = 1~3; transition metal "M" = Sc, Ti, Zr, Hf, V, Nb, Mo, and so on; "X" = carbon and/or nitrogen and "$T_x$" is the surface terminations like hydroxyl, oxygen, or fluorine[41]), they have useful optical applications for saturable absorbers, photodetectors, and modulators[133-136].

Metal-halide perovskites are representative organic–inorganic hybrid perovskites that have attracted tremendous research interest over the past decade for their promising optoelectronic attributes[137] for applications in photovoltaics, lasers, LEDs, photodetectors, and nonlinear optics[47,59,138-148]. They typically have corner-sharing $BX_6^{4-}$ octahedra, where "B" stands for a divalent metal cation like $Pb^{2+}$, $Sn^{2+}$ or $Ge^{2+}$, and "X" denotes a monovalent halide anion[143]. Inspired by advents in 2D materials community, recent study shows that these materials can be also made to molecularly thin versions like 2D materials[149]. With varying dimensions and layer numbers $n$, quasi-2D metal-halide perovskites have different electronic properties, excitonic coupling strength, and varying degrees of quantum-and dielectric-confinement effects for integration with nanophotonic devices[150].

## 2.2 Photonic structures for coupling with 2D materials

In nanophotonics, judiciously designed optical structures are essential for a vast number of applications[32]. For instance, dielectric waveguides are the crucial building block in photonic integrated circuits[6,14]. When 2D materials are transferred to a photonic waveguide or an optical fiber, it can evanescently couple with the guided electromagnetic modes inside to alter its effective mode index, light propagation, and dispersion[1].

Optical cavities can spatially confine light into small volumes[151-153] to boost light-matter interaction strength for nonlinear[153-157], lasing[158], and sensing applications[25,159,160] when integrated with 2D materials. The synergy of temporal accumulation and spatial confinement of photons can drastically enhance the efficiency of encapsulated 2D materials to the optical resonators.

Photonic crystals are periodic nanostructures with dimension features similar to light wavelength[161]. Under the periodic refractive index perturbations, photonic bandgap emerges[162]. They can also be applied as optical cavities for light enhancement and light guiding structure by intentionally introducing point and line defects respectively[162,163].

In contrast, metasurfaces are engineered optical scatterers array with subwavelength dimension that can achieve powerful control over the fundamental attributes of light[28,113,164-177], such as the phase, wavevector, amplitude, and frequency[178-185], with wide applications such as meta-lenses[186,187], high-efficiency holograms[188-190], color display[27,190-192], ultrathin cloak[193-195], healthcare sensors[196-199], LiDAR[200-202], functional meta-waveguides[14,203-222], and nonlinear applications[223-225], to name a few.

## 3. Applications on 2D photonics

Armed with the diverse 2D van der Waals materials building blocks and a large library of photonic structures, massive optical and optoelectronic applications can be prototyped. The recent advents and opportunities are outlined below based on the primal photonic structures that coupled with 2D materials.

### 3.1 Optical waveguides

As the essential building block of photonic integrated circuit[6], photonic waveguides are physical structures to guide the propagation of electromagnetic waves[14]. By transferring 2D materials such as graphene[34,226], the effective mode index of the underlying dielectric bus waveguide can be controlled via the structure engineering of the waveguide[227], as well as via optical pumping[228], electrical gating[94], strain[229], or thermal tuning[230,231] of the graphene monolayer for optical amplitude[232,233] or phase modulators[105,234]. The giant optical nonlinearity in graphene can also be harvested using graphene-laminated waveguide structures for gate-controlled optical nonlinearity[95,235]. The graphene-integrated waveguides can be leveraged as broadband photodetectors for harvesting guided light signals as well[81,93,109,236-238]. For other 2D materials such as TMD and black phosphorous, similar applications are also developed while with different operation wavelengths[239-242], and they can be more promising for active optical applications due to the varying bandgap values of these 2D materials[35,53,121].

Besides, graphene is also an excellent platform to exploit surface plasmon polaritons[98,226,243]. Plasmonic graphene modulators may thus have much compact footprint and fast operation speed albeit higher optical loss[244-246]. Low-loss surface plasmons can be supported in graphene/h-BN heterostructures with favorable performance at terahertz regions[97,247-253]. In contrast, other 2D materials such as TMD are more favorable to exploit nanophotonic polaritons and excitons[254-256]. Graphene and other 2D materials can also serve to add reconfigurability to control the coupling[206,257,258] and dispersion[259-261] of the optical waveguides to permit chip-scale optical signal processing. Applications such as plasmonic sensors[262-264], transformation optics[265], nano-imaging[96,266,267], gain-modulated lasing applications[268,269] can be prototyped as well.

### 3.2 Micro-cavities

Optical cavities with varying formats are ideal for optical applications that require optical field enhancement and/or strong light-matter interactions[270]. Photons can be confined into very small volumes by resonance and recirculation[271]. For instance, Fabry-Perot microcavities fabricated by depositing Bragg reflectors, engineered photonic crystals, and whisper-gallery resonators such as micro-sphere, micro-toroid, and micro-ring resonators with varying quality (Q) factor from thousands to billions[272,273]. The 2D materials such as TMD and black phosphorous can be transferred to or incapsulate into prefabricated optical cavities to harvest strong light enhancement for nano-lasers[274-279], optical amplification[280,281], sensing[71,282], and photodetection[283-285] to make most of the atomically thin semiconductor materials with optical gain[286].

Another exemplary application direction of 2D materials-coupled optical cavities is nonlinear optics[287,288]. Engineered giant light field enhancement and nonlinear filed overlap, the nonlinear optical response in graphene, TMD, quasi-2D perovskites, black phosphorous, and so on can be boosted in the optical nanocavities[289-293] for modulation[294-296], switching[297,298], and harmonic generation[299-301], to name a few. Alternatively, the 2D material nano-sheets can also be directly patterned into nano-resonators, frequency conversion[302].

## 3.3 Metasurfaces

Metasurfaces with artificially engineered optical nanoantennas have showcased unprecedented degrees of freedom in controlling massive fundamental light attributes[165,183]. By judiciously altering the material and structure design of the subwavelength nanostructures, the reflected or refracted electromagnetic wave can be flexibly tailored[303-308]. As a 2D counter part of metamaterials[309-311], intriguing nanophotonic phenomena and applications can be hatched by combining the 2D nanomaterials[312,313].

The 2D materials as an ultrathin nano-sheet can be transferred on top of metasurface to permit electrically or optically reconfigurable or programmable meta-devices for direct polarization detectors[314], right routing[76], beam steering[315], lasing[316], and sensing[317-319]. The metasurface structure can also enable enhanced light-2D material applications for spontaneous control over harmonic signal generation and beam control[169,320]. At the same time, the 2D materials can be also patterned into atomically thin metasurfaces for tunable planner optics[321-323], light sources[324,325], beam splitters[326], and so on.

## 4. Outlook

Despite that high-quality 2D materials can be produced via mechanical exfoliation[37], they are typically small in size and can hardly meet future industrial practical applications[105,327]. In terms of further commercialization of 2D-materials-based photonic and optoelectronic devices[36,328], wafer-scale scalable and reproducible 2D material growth and transfer are required[329,330]. Considering the yield and quality, chemical vapor deposition (CVD)-based approaches may permits potentially higher throughput and lower cost compared to molecular beam epitaxy[331,332]. Nevertheless, besides graphene, the direct CVD synthesis of continuous and uniform monolayer 2D materials such as TMD and h-BN is still challenging[331,333-337]. To address this trade-off in monolayer controllability and large-scale film uniformity of high-quality 2D materials, geometrically confined 2D material growth with pre-defined growth pockets can be considered[338]. The ultimate goal might be the direct synthesis of high-quality 2D monolayers or 2D heterostructures, as a transfer-free method, on top of prefabricated photonic structures[339,340]. For transfer of 2D materials, roll-to-roll scalable transfer[341-343] or robots-based precise automatic 2D flake transfer[344,345] will be favorable for industrial production[36].

Recently, several emerging 2D materials with exotic attributes have further opportunities[46,143,149], such as 2D van der Waals materials with vanished interlayer coupling for scalable nonlinear optical applications[346], quasi-2D perovskites with high crystallinity[45,48,149], and ferroelectricity and ferromagnetism in atomically thin van der Waals layered materials[347-350]. Confined growth-enabled large-scale integrated photonic devices[338] is also promising for future industrialized 2D integrated photonic circuits[14,339].

Besides 2D materials that have Minuscule thickness limiting their interaction length with electromagnetic waves[1], recent advances on remote epitaxy[351-353] and van der Waals epitaxy[354-357] also permits various three-dimensional (3D) freestanding nanomembranes[1,355,358-362], as well as other 3D layer exfoliation techniques by epitaxial chemical lift-off, mechanical exfoliation or laser lift-off[1,85,363-369]. These thin films are also made ultrathin with artificially defined van der Waals interfaces for photonic van der Waals integration[1,58,59]. These 3D nanomembranes can preserve their optical attributes almost independent from film thickness[85,369] for versatile light-emitting[370], sensing[371], smart embedded optoelectronic processors[372], light enhancement[373-375], new integrated photonics platforms[376], solar cells[377], flexible LEDs[378,379], and other novel hetero-integrated devices[85,380,381]. Nanomembranes mass production could be also enabled with reduced cost[378,382-385].

By combining the 2D and 3D van der Waals material building blocks, vast playgrounds also unfolds for 2D-materials-based photonic integrated circuits[6,7,32,82,386], exotic nanophotonic polaritons in hybrid heterostructures[5,73,74], flexible, wearable and implantable optoelectronic biosensors[26,64,196,262], and vertical 3D integrated circuity and systems[372,387-390].